\documentclass[twocolumn,showpacs,preprintnumbers,amsmath,amssymb,prl,aps]{revtex4}

\usepackage{epsfig}
\usepackage{graphicx}
\usepackage{dcolumn}
\usepackage{bm}

\begin{document}

\title{Manifestation of ground state anomaly in the core level spectra
of Ca and Sr in Ca$_{1-x}$Sr$_x$RuO$_3$}

\author{Ravi Shankar Singh and Kalobaran Maiti}

\affiliation{Department of Condensed Matter Physics and Materials
Science, Tata Institute of Fundamental Research, Homi Bhabha Road,
Colaba, Mumbai - 400005, INDIA}

\date{\today}

\begin{abstract}

We investigate the evolution of Ca 2$p$ and Sr 3$d$ core level
spectra in Ca$_{1-x}$Sr$_x$RuO$_3$ using photoemission
spectroscopy. Core level spectra in this system exhibit multiple
features and unusual evolution with the composition and
temperatures. Analysis of the core level spectra in conjunction
with the band structure results reveal novel final state effects
due to different core hole screening channels. Changes in the core
level spectra suggest significant modification in Ca-O covalency
in Ca dominated samples, which gradually reduces with the increase
in Sr content and becomes minimum in SrRuO$_3$. This study thus,
provides a direct evidence of the role of cation-oxygen covalency
in the ground state properties of these novel materials.

\end{abstract}


\maketitle

Ruthenates have drawn significant attention in the recent time due
to many interesting properties such as superconductivity
\cite{sr2ruo4}, non-Fermi liquid behavior \cite{nfl,klein},
unusual magnetic ground states \cite{nfl,klein,rss,cao} {\it etc.}
observed in these materials. SrRuO$_3$, a perovskite compound
exhibits ferromagnetic long range order ($T_C$~=~160~K) despite
highly extended 4$d$ character of the valence electrons
\cite{rss,cao}. Interestingly, CaRuO$_3$, an isostructural
compound exhibit similar magnetic moment at high temperatures as
that observed in SrRuO$_3$, however, no long-range order is
observed down to the lowest temperature studied
\cite{nfl,klein,rss,cao}. These investigations predict a non-Fermi
liquid ground state in CaRuO$_3$ in contrast to the Fermi-liquid
behavior observed in SrRuO$_3$ \cite{nfl,klein}. Both SrRuO$_3$
and CaRuO$_3$ form in an orthorhombic perovskite structure
(ABO$_3$ - type) \cite{ramarao,nakatsugawa}. It is believed that
the A cation (Sr/Ca) sites help to form the typical building block
of this structure and the RuO$_6$ octahedra connected by corner
sharing essentially determines the electronic properties. The
Ru-O-Ru bond angles are somewhat different in these compounds
(165$^\circ$ in SrRuO$_3$ and 150$^\circ$ in CaRuO$_3$) suggesting
an enhancement in effective electron correlation strength.
However, a recent experimental study shows that such effects are
significantly weak as expected for a highly extended 4$d$
transition metal oxides \cite{ravi} and thus, the experimental
observation of different ground state properties still remains a
puzzle.

Photoemission spectroscopy has widely been used to study these
systems. However, the electronic states corresponding to A-site
cations have often been neglected due to the absence of
significant contribution from these elements in the vicinity of
the Fermi level. In this study, we investigate the evolution of
the core level spectra associated to the A-site cations. High
quality polycrystalline samples (large grain size achieved by long
sintering at the preparation temperature) were prepared by solid
state reaction method using ultra-high purity ingredients and
characterized by $x$-ray diffraction (XRD) patterns and magnetic
measurements as described elsewhere \cite{rss,ravi}. Sharp XRD
patterns reveal pure GdFeO$_3$ structure with similar lattice
constants as observed for single crystalline samples \cite{cao}.
Photoemission measurements were performed on {\it in situ}
(4$\times$10$^{-11}$~torr) scraped samples using SES2002 Scienta
analyzer. Reproducibility and cleanliness of the sample surface
was confirmed after each trial of scraping.

\begin{figure}
\vspace{-2ex}
 \centerline{\epsfysize=4.5in \epsffile{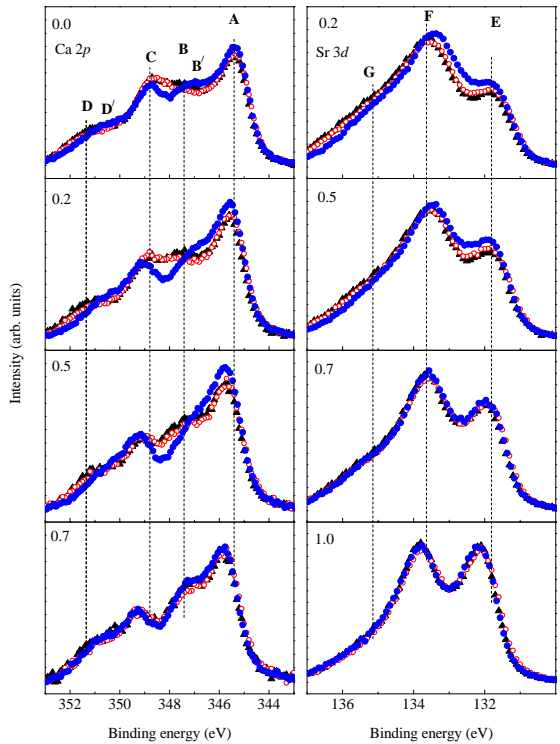}}
\vspace{-4ex}
 \caption{Ca 2$p$ (left column) and Sr 3$d$ (right column) core level
spectra in Ca$_{1-x}$Sr$_x$RuO$_3$ for different values of $x$.
Triangles, open circles and solid circles represent the spectra
collected using Mg $K\alpha$ radiations at 300 K, Al $K\alpha$
radiations at 300 K and Al $K\alpha$ radiations at 20 K,
respectively.}
 \vspace{-2ex}
\end{figure}

In Fig.~1, we show Ca 2$p$ and Sr 3$d$ core level spectra for
$x$~=~0.0, 0.2, 0.5, 0.7 and 1.0 in Ca$_{1-x}$Sr$_x$RuO$_3$. All
the spectra are normalized by integrated intensity and exhibit
multiple features in contrast to the spin orbit split two peak
structure expected. This is unusual as Ca$^{2+}$ and Sr$^{2+}$ are
believed to be highly ionic in nature with the unoccupied $d$
bands appearing significantly above the Fermi level. There are
four distinct features marked by A, B, C and D in all the Ca 2$p$
spectra shown in the left column of the figure. Despite large spin
orbit splitting ($\sim$~3.5~eV) for the Ca 2$p$ signals,
experiments on several sets of samples in different experimental
setups and high energy resolution (300~meV) do not exhibit any
significant difference in the spectral features. A reduction in
excitation energy (from Al~$K\alpha$ (open circles) to
Mg~$K\alpha$ (triangles)) leads to a small increase in intensity
of the features B and D. Since such a change in photon energy
increases the surface sensitivity of the technique due to the
reduction of the photoelectron kinetic energies, such spectral
modification is often considered as the signature to identify the
bulk and surface related features \cite{csvoepl,csvoprb}. However,
several intriguing effects observed on variation of different
parameters as described below, cannot be explained within this
scenario.

\begin{figure}
\vspace{-2ex}
 \centerline{\epsfysize=4.5in \epsffile{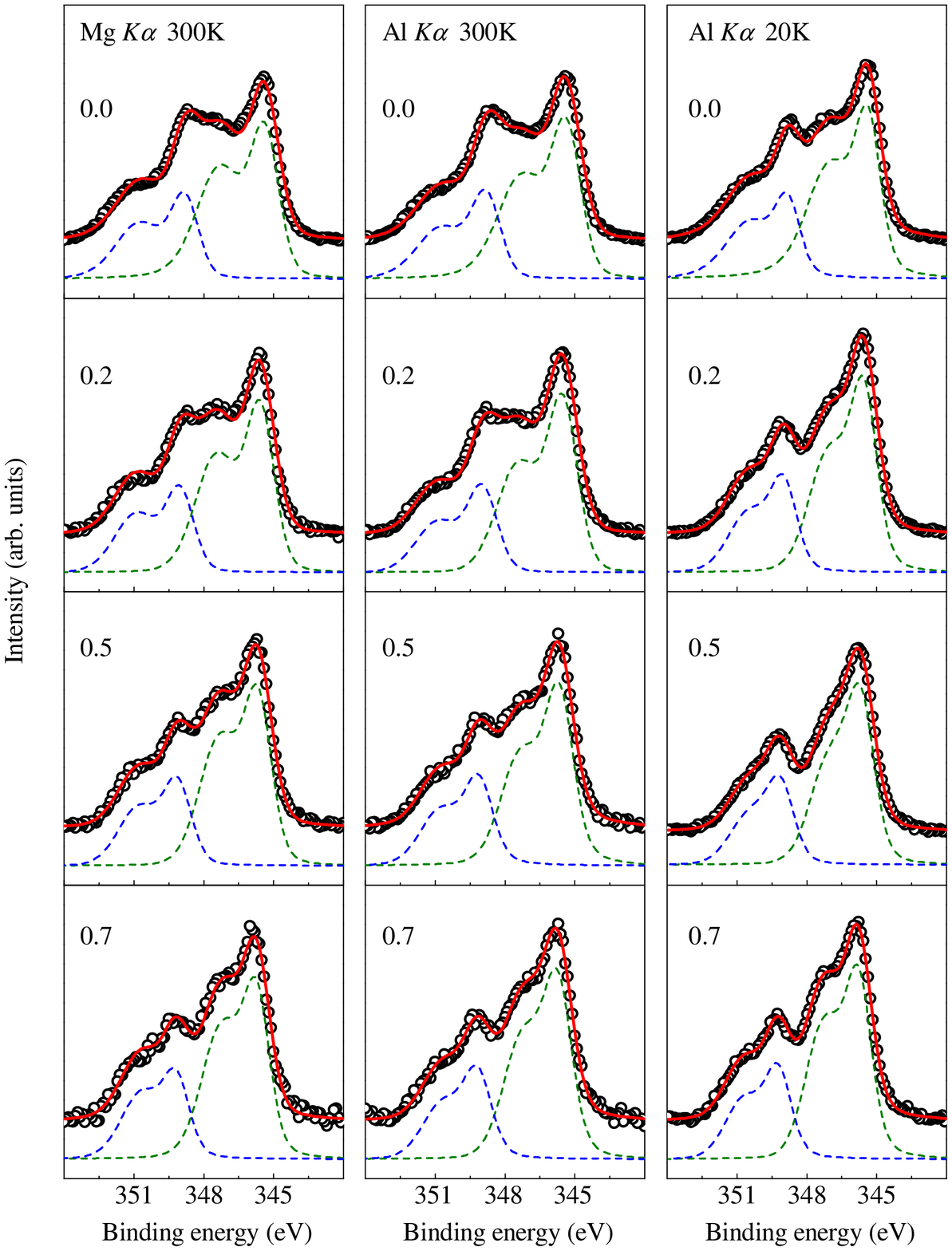}}
\vspace{-4ex}
 \caption{Experimental (circles) and simulated (solid lines) Ca 2$p$
spectra for different $x$ in Ca$_{1-x}$Sr$_x$RuO$_3$. The dashed
lines represent the 2$p_{3/2}$ and 2$p_{1/2}$ features delineated
from the experimental spectra.}
 \vspace{-2ex}
\end{figure}

We show the Al~$K\alpha$ spectra at 20~K by solid circles in the
same figure. While features A and C remains at the same energies,
a shift of the features B and D to lower binding energies is
clearly evident as shown by B$^\prime$ and D$^\prime$ in the
figure. This reveals that the changes in the features A and C
separated by about 3.5~eV are very similar, and the features B and
D are connected together with an energy separation of about
3.5~eV. Thus, the features A and B (B$^\prime$) can be attributed
to the photoemission signal due to the excitation of Ca 2$p_{3/2}$
electrons, and the features C, D (D$^\prime$) appear due to
2$p_{1/2}$ electronic excitations. The spectral modifications are
found to be significantly large in CaRuO$_3$. Interestingly, the
peak position of the features A and C gradually shifts to higher
binding energies with the increase in $x$, while the features B
and D appears almost at same binding energies across the series.
In addition, the spectral modification with temperature gradually
becomes insignificant with the increase in $x$.

\begin{figure}
\vspace{-2ex}
 \centerline{\epsfysize=4.5in \epsffile{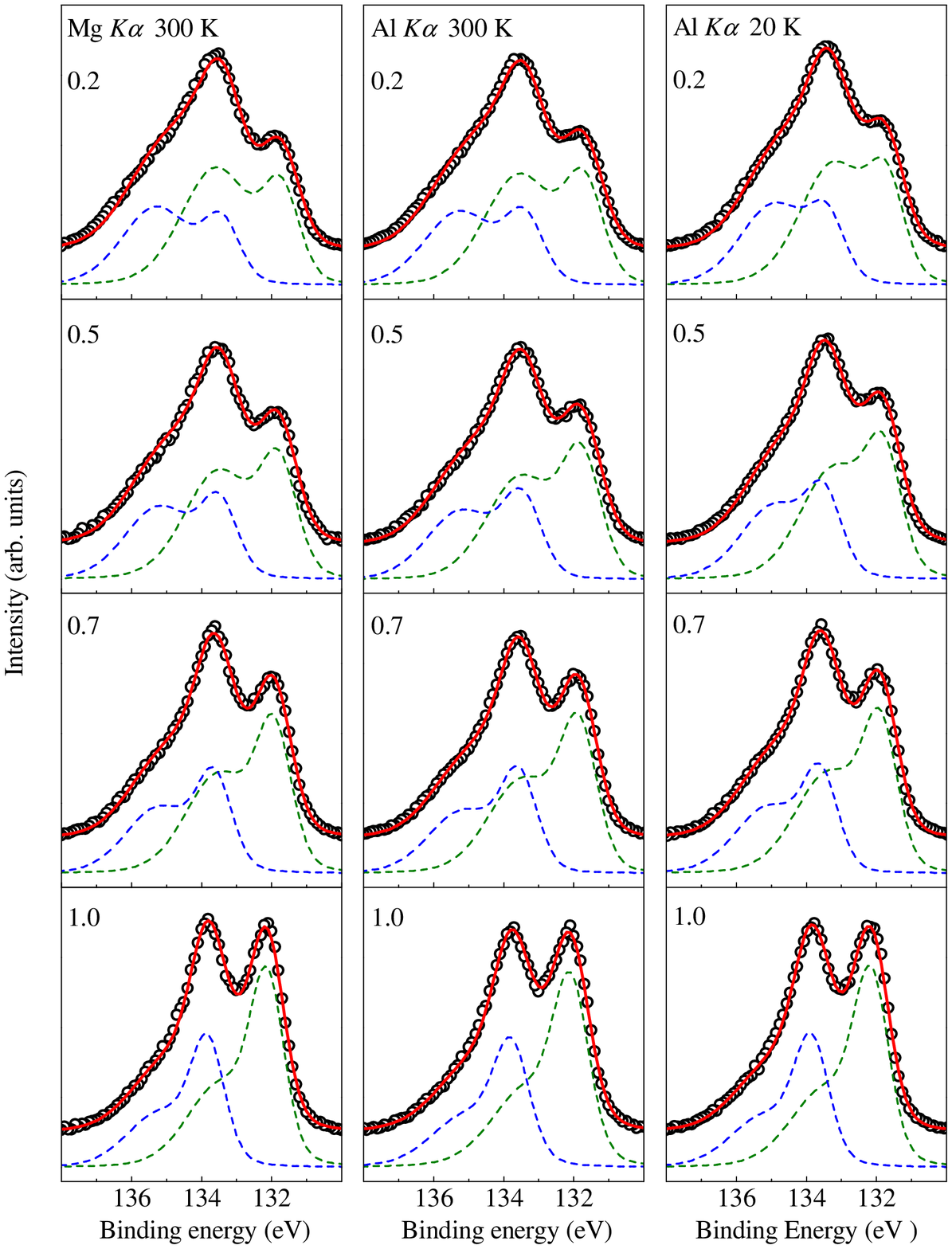}}
\vspace{-4ex}
 \caption{Experimental (circles) and simulated (solid lines) Sr 3$d$
spectra for different $x$ in Ca$_{1-x}$Sr$_x$RuO$_3$. The dashed
lines represent the 3$d_{5/2}$ and 3$d_{3/2}$ features delineated
from the experimental spectra.}
 \vspace{-2ex}
\end{figure}

We now turn to the Sr 3$d$ core level spectra. Although, a
distinct set of spin orbit split 3$d$ signal is visible in the Sr
3$d$ spectra of SrRuO$_3$ ($x$~=~1.0), the ratio of the intensity
of 3$d_{5/2}$ and 3$d_{3/2}$ does not follow the expected
multiplicity ratio of 3:2. All the spectra in the intermediate
compositions exhibit three distinct features as shown by E, F and
G in the figure. Most interestingly, the feature F appearing at
about 133.6 eV binding energy exhibit the largest intensity, in
sharp contrast to the expectation of lower intensity corresponding
to the spin orbit split 3$d_{3/2}$ signal compared to the
3$d_{5/2}$ signal, E appearing at about 131.8~eV. The change in
photon energy from Al~$K\alpha$ to Mg~$K\alpha$ does not exhibit
significant modifications in the spectral features. However, the
decrease in temperature leads to a shift of the feature F towards
lower binding energies in Ca$_{0.8}$Sr$_{0.2}$RuO$_3$. With the
increase in $x$, this shift becomes gradually insignificant as
also observed in Ca 2$p$ spectra. The feature E shifts towards
higher binding energies with the increase in $x$. Interestingly,
the change in photon energy and temperature has no influence in
the Sr 3$d$ spectra in SrRuO$_3$.

In order to investigate the spectral changes in more detail, we
have simulated all the core level spectra using two sets of
Lorentzians (representing lifetime broadening of the photoholes),
where each set represents a spin orbit split two signals. These
Lorentzians are convoluted with a Gaussian to consider resolution
broadening and other solid state effects. In order to reduce the
uncertainty in the fitting procedure, we have fixed the spin-orbit
splitting to 3.47~eV for Ca 2$p$ and 1.7~eV for Sr 3$d$ spectra
found in CaRuO$_3$ and SrRuO$_3$, respectively. The fits are
carried out using least square error method by varying the energy
separation and the relative intensity of two sets of features.

The simulated spectra for all the Ca 2$p$ spectra are shown by
solid lines in Fig.~2 by overlapping over the experimental
spectra. All the fits exhibit a beautiful representation of the
experimental spectra. Clearly, 2$p_{3/2}$ and 2$p_{1/2}$ features
contains at least two features in each case. The energy separation
between the features in CaRuO$_3$ is found (error bar = 0.02~eV)
to be 1.9~eV, which reduces to 1.6~eV at 20~K. The energy
separation gradually reduces with the increase in $x$ (1.88~eV,
1.65~eV and 1.5~eV for $x$ = 0.2, 0.5 and 0.7, respectively).
However, the low temperature spectra do not exhibit such large
change (1.5~eV, 1.4~eV and 1.45~eV for $x$ = 0.2, 0.5 and 0.7,
respectively). The most notable point is that the difference
between room temperature and low temperature spectra gradually
reduces with the increase in $x$. The simulated Sr 3$d$ spectra
are shown by solid lines overlapped over the experimental ones in
Fig.~3. Anomalously high intensity observed for the feature F in
the experimental spectra (see Fig.~1) could be simulated exactly
considering two peak structure for the photoemission signal
corresponding to each spin-orbit split features. The intensity of
the higher binding energy feature is found to be significantly
high compared to other one in the Ca-dominated compositions. This
intensity reduces drastically with the increase in Sr-content
across the series. The energy separation between the two features
is found to be 1.85~eV at room temperature for $x$~=~0.2. With the
increase in $x$ this separation reduces gradually (1.7~eV, 1.6~eV
and 1.3~eV for $x$ = 0.5, 0.7 and 1.0, respectively). The
reduction in temperature leads to a decrease in the separation of
the two features to 1.5~eV, 1.4~eV, 1.4~eV and 1.3~eV for $x$ =
0.5, 0.7 and 1.0, respectively. Thus, the influence of temperature
on the energy separation between the two features in each
spin-orbit split signal gradually reduces with the increase in
Sr-content and becomes invisible in SrRuO$_3$.

It is now clear that the two peak structure in the Ca 2$p$ and Sr
3$d$ core level spectra cannot be attributed to the differences in
Madelung potential at A-sites present in the structure and/or
surface-bulk related differences. In particular, various
structural analysis also suggest that all the A-site cations are
equivalent even if the structure is significantly distorted from
the cubic perovskite structure. It is to note here that various
binary and ternary compounds involving rare-earths at the A-site
often exhibit additional features in the core level spectra. For
example, La 3$d$ spectra in La$_2$O$_3$, La$M_3$ ($M$ is a
monovalent element such as F, Br, Pd {\it etc.}
\cite{hillebrecht,crecelius,signorelli}) exhibit distinct two
peaks due to different screening channels in the photoemission
final states. Thus, the two peak structures in the Ca 2$p$ and Sr
3$d$ core level spectra in these compounds can be attributed to
differently screened final states.

The ground state wave function can be expressed as a linear
combination of electronic states having different $d$-band
occupancy such as $\mid d^0>$, $\mid d^1L>$ {\it etc.}, where $L$
represents a hole in the oxygen band. The $d$ bands corresponding
to Ca (3$d$) and Sr (4$d$) are almost empty and appear
significantly above the Fermi level ($>$~4~eV). Thus, the charge
transfer energy, the energy required to transfer an electron from
the O 2$p$ bands to the $d$-bands is expected to be significantly
high. Therefore, the contributions from the charge transferred
states ($\mid d^1L>$, $\mid d^2L^2>$ {\it etc.}) will be
energetically less favorable and the initial state will be
primarily contributed by $\mid d^0>$ electronic configuration. In
the final state, core holes created by the photoemission enhance
the local positive charge density. Hence, the core holes are often
screened by the electrons transferred from the ligand bands and
are energetically most favorable. The photoemission signals
corresponding to these final states are known as {\em well
screened} or {\em main} peak and appear at lower binding energies.
The feature corresponding to the core holes without any screening
by such charge transfer are known as {\em poorly screened} or {\em
satellite} features. The energy separation and the relative
intensity of the main and satellite features depend on the charge
transfer energy and the O 2$p$~-~A-site $d$ hybridization. It is
to note here that if this hybridization is zero, the well screened
feature will be insignificant.

\begin{figure}
\vspace{-2ex}
 \centerline{\epsfysize=4.5in \epsffile{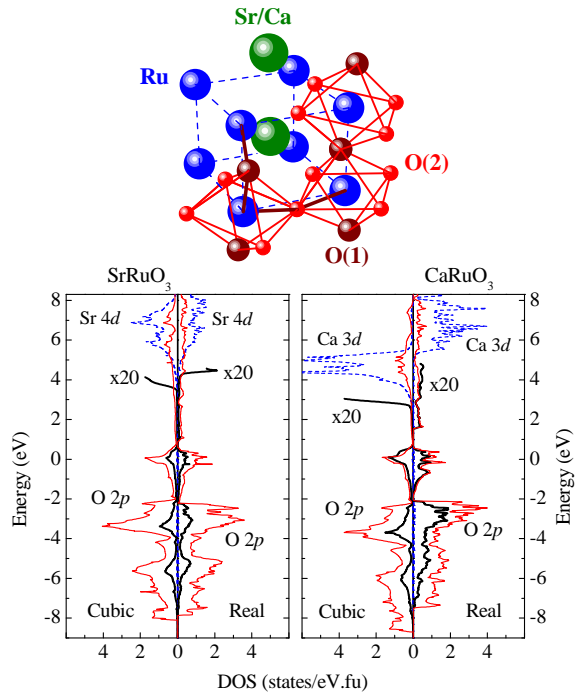}}
\vspace{-4ex}
 \caption{Crystal structure of SrRuO$_3$ and CaRuO$_3$. Band
structure results for SrRuO$_3$ and CaRuO$_3$ in the real and
equivalent cubic structures are shown in lower left and right
panels, respectively. O 2$p$ and Sr 4$d$/Ca 3$d$ PDOS are
represented by thin solid lines and dashed lines, respectively.
The thick solid line represent the Sr 4$d$/Ca 3$d$ PDOS multiplied
by 20 to amplify the low energy part.}
 \vspace{-2ex}
\end{figure}

In order to investigate the Ca-O and Sr-O hybridization, we have
calculated the electronic band structure of this system using full
potential linearized augmented plane wave method within the local
spin density approximations ({\scriptsize WIEN2K}
software\cite{wien}). The calculated results for Sr 4$d$ and O
2$p$ partial density of states (PDOS) in the real crystal
structure and in the equivalent cubic structure (unit cell volume
kept fixed) of SrRuO$_3$ are shown in the left panel of Fig.~4. Sr
4$d$ contributions essentially appears above 3.6~eV in the cubic
structure as shown by scaling the Sr 4$d$ PDOS by 20 times. A
significant contribution from O 2$p$ PDOS appears in this energy
range. Interestingly, the Sr 4$d$ band edge shifts to 4.4~eV in
the real structure along with a significant modification in the O
2$p$ PDOS. Such changes indicate finite Sr-O covalency forming
bonding and antibonding bands, which leads to a GdFeO$_3$ kind of
distortion by moving O(1) towards A-site as shown in
Fig.~4.\cite{kbm} The bonding band with primarily O 2$p$ character
appears below the Fermi level and the antibonding band having
primarily Sr 4$d$ character appears above the Fermi level. The
calculated results in CaRuO$_3$ are shown in the right panel of
Fig.~4. The Ca 3$d$ PDOS exhibit the lower energy edge with
significant intensity at 2.6~eV as expected for 3$d$ orbitals
compared to 4$d$ orbitals in SrRuO$_3$. Most strikingly, the 3$d$
band shifts to much higher energies (band edge at 5.5~eV) compared
to that observed in the 4$d$ band in SrRuO$_3$. Such a strong
modification is unusual and clearly reveal strong Ca-O covalency
effects due to the large overlap of the O 2$p$ and Ca 3$d$ bands
similar to that in vanadates \cite{andersen}.

These results establish that the hybridization of the $d$ bands
associated to A-site elements with O 2$p$ bands is significant.
The higher energy of the Ca 3$d$ band compared to Sr 4$d$
conduction band suggests that the charge transfer energy for the
electrons from O 2$p$ bands to A-site $d$ bands is significantly
larger in CaRuO$_3$ than that in SrRuO$_3$. This explains the
observation of larger energy separation between the well screened
and poorly screened features in the core level spectra, and the
gradual reduction of this separation with the increase in
Sr-content.

The GdFeO$_3$ distortion observed in this system, reduces the
Ru-O-Ru bond angles significantly from 180$^\circ$. Subsequently,
the RuO$_6$ octahedra also becomes distorted leading to different
O-O couplings. Thus, the O 2$p$ bandwidth reduces significantly as
evident in the calculated results for real structure in comparison
to that in the equivalent cubic structure. Such narrowing
significantly affects the delocalization of holes created in the O
2$p$ band due to charge transfer. This is manifested by an
increase in intensity of the poorly screened feature compared to
the well screened feature intensity towards CaRuO$_3$ end of
Ca$_{1-x}$Sr$_x$RuO$_3$ in both Ca 2$p$ and Sr 3$d$ spectra (see
Figs. 2 and 3).

We now turn to the question of temperature evolution in the core
level spectra. The peak position of the satellite feature shifts
(0.3~eV in CaRuO$_3$ and gradually reduces with increase in $x$)
to lower binding energies. This can qualitatively be explained as
follows. It is now clear that the strong A-O covalency leads to a
GdFeO$_3$ distortion in the crystal structure \cite{kbm,andersen}.
While large covalency increases the separation between the
unoccupied $d$ band (antibonding) and occupied O 2$p$ band
(bonding), the center of mass of the bonding band is observed to
shift towards the Fermi level (see Fig.~4). The decrease in
temperature leads to a compression of the crystal lattice. While
such compression leads to an enhancement in the bandwidth due to
reduction in bond lengths (a reduction in charge transfer energy),
it often introduces a larger degree of distortion as observed in
the bulk properties due to application of external pressure on
SrRuO$_3$ \cite{pressure}. Thus, the effective potential at the
A-sites will be modified with the decrease in temperature leading
to a reduction in binding energy of the poorly screened feature in
addition to a decrease in the main peak-satellite separation due
to the reduction in charge transfer energy. It is clear that the
electronic structure in CaRuO$_3$ is significantly influenced by
the change in temperature, however, that in SrRuO$_3$ is
relatively insensitive. While it is natural to draw an
interconnection of such temperature induced changes in electronic
structure to the non Fermi liquid behavior in CaRuO$_3$, further
studies are required to understand the temperature evolutions in
great detail.

In summary, we have investigated the evolution of the core level
spectra associated to the A-site in ABO$_3$ structure as a
function temperature and composition in Ca$_{1-x}$Sr$_x$RuO$_3$.
Photoemission results exhibit multiple structures in the core
level spectra. Analysis of the spectra reveal strong influence of
core hole screening in the final states, a novel effect associated
to A-sites in the ABO$_3$ structure not studied so far. Evolution
of the core level spectra with temperature and composition suggest
that the distortion in the crystal structure and its change with
temperature possibly play the key role in determining
significantly different ground state properties in these
interesting materials.

\end{document}